\documentclass[aps,prapp,reprint]{revtex4-2}
\usepackage{amsmath}
\usepackage{amssymb}
\usepackage{mathrsfs}
\usepackage{graphicx}
\usepackage{booktabs}  
\usepackage{multirow}  
\usepackage[colorlinks,citecolor=blue, linkcolor=blue,hyperindex,CJKbookmarks,]{hyperref}

\begin{document}

\title{Multiterminal Nonreciprocal Routing in an Optomechanical Plaquette \\ via Synthetic Magnetism}
\author{Zhi-Xiang Tang}
\affiliation{Key Laboratory of Low-Dimensional Quantum Structures and Quantum Control of
Ministry of Education, Key Laboratory for Matter Microstructure and Function
of Hunan Province, Department of Physics and Synergetic Innovation Center
for Quantum Effects and Applications, Hunan Normal University, Changsha
410081, China}
\author{Xun-Wei Xu}
\email{xwxu@hunnu.edu.cn}
\affiliation{Key Laboratory of Low-Dimensional Quantum Structures and Quantum Control of
Ministry of Education, Key Laboratory for Matter Microstructure and Function
of Hunan Province, Department of Physics and Synergetic Innovation Center
for Quantum Effects and Applications, Hunan Normal University, Changsha
410081, China}

\date{\today}

\begin{abstract}
Optomechanical systems with parametric coupling between optical (photon) and
mechanical (phonon) modes provide a useful platform to realize various magnetic-free nonreciprocal devices, such as isolators, circulators, and directional amplifiers.
However, nonreciprocal router with multiaccess channels has not been extensively studied yet.
Here, we propose a nonreciprocal router with one transmitter, one receiver, and two output terminals, based on an optomechanical plaquette composing of two optical modes and two mechanical modes.
The time-reversal symmetry of the system is broken via synthetic magnetism induced by driving the two optical modes with phase-correlated laser fields.
The prerequisites for nonreciprocal routing are obtained both analytically and numerically, and the robustness of the nonreciprocity is demonstrated numerically.
Multiterminal nonreciprocal router in optomechanical plaquette provides a useful quantum node for development of quantum network information security and realization of quantum secure communication.
\end{abstract}

\maketitle

\section{Introduction}

In a linear magnetic-free system, the transport of light is governed by the Lorentz reciprocity theorem~\cite{Born1999book,Potton_2004}.
However, nonreciprocal optical devices are the basic building blocks for information processing and sensing~\cite{Jalas2013NaPho,Caloz2018PRAPP}, so we need to break the time-reversal symmetry of the system.
Recently, in order to satisfy the demand of on-chip integrated information processing, the research of magnetic-free optical nonreciprocity scheme has attracted a lot of attention.
As one of the most promising magnetic-free nonreciprocal schemes, optomechanically induced nonreciprocity has attracted much interest in the past decade~\cite{Aspelmeyer2014RMP,Verhagen2017NatPh,Barzanjeh2022NatPh}, and various nonreciprocal mechanisms based on optomechanical interactions are proposed theoretically and demonstrated experimentally, such as direction-dependent optomechanical nonlinearity~\cite{manipatruni2009PRL, wang2015SR,XuXW2018PRA,SongLN2019PRA}, microring with unidirectional pumping~\cite{Hafezi2012OExpr,Shen2018NatCo,Ruesink2018NatCo,Dong2016NaPho,Ruesink2016NatCo,XuXW20PRJ,TangZX2023PRAPP}, stimulated Brillouin scattering~\cite{Dong2015NatCo,Kim2015NatPh}, dynamically encircling an exceptional point~\cite{XuH2016Natur,Zhang2022Nanop,LongD2022PRA}, Sagnac effect in spinning resonator~\cite{LiBJ19PRJ,JiaoYF2020PRL,LiWA2020PRA,LiBJ2021PRA,ZhangDW2021PRA,JiaoYF2022PRAPP,PengM2023PRA,Shang2022Photo}, and quantum interference based on synthetic magnetism~\cite{Schmidt2015Optic,Metelmann2015PRX,XuXW2015PRA}.

Optomechanical plaquette formed from the synthetic dimension created by the cycle coupling of optical and mechanical modes provides a ideal platform for nonreciprocity via synthetic flux threading the plaquette.
Based on this plaquette, various nonreciprocal devices have been proposed theoretically, including optical isolator and circulator~\cite{XuXW2015PRA,TianL2017PRA,Miri2017PRAPP,XuXW2020PRAPP,Yan2019FrPhy}, nonreciprocal frequency converter~\cite{XuXW2016PRA}, directional optical amplifier~\cite{LiY2017OExpr,Malz2018PRL,JiangC2018PRA,LanYT2022OptL}, nonreciprocal ground-state cooling~\cite{Habraken2012NJPh,XuH2019Natur,LaiDG2020PRA}, asymmetric optomechanical entanglement~\cite{LiuJX2023SCPMA}, and such structure has been extended to optomechanical arrays to realize non-reciprocal control of phonon flow~\cite{Seif2018NatCo,Denis2020PRL,Barzanjeh2018PRL} and simulate gauge fields in many-body physics~\cite{Schmidt2015NJP,Peano2015PRX,Sanavio2020PRB}.
Similar mechanisms have been realized for optical photons in optomechanical crystals~\cite{Fang2017NatPh,Mathew2020NatNa,Pino2022Natur} and microresonators~\cite{ChenY2021PRL}, and microwave photons in superconducting circuits~\cite{Peterson2017PRX,Bernier2017NatCo,Barzanjeh2017NatCo,Mercier2019PRAPP,Mercier2020PRL}.

Router for controlling the path of signals, is a key element in constructing network, and has been carried out
in multi-waveguides connected by atoms~\cite{GU20171,ZhouL2013PRL,LuJ2014PRA,Gonzalez2016PRA,YangDC2018PRA,ZhouJ2023PRA}, QED systems~\cite{Aoki2009PRL,XiaK2013PRX,JiaW2019PRA,WangZL2021PRL}, or optomechanical systems~\cite{Agarwal2012PRA,LiX2016NatSR,XuXW2017PRA}, and one bus-waveguide side coupled to numerous quantum emitters~\cite{Hoi2011PRL,WuJN2022PRAPP,Yan2014NatSR}.
For the further development of quantum network information security and the realization of quantum secure communication, 
nonreciprocal router that steers the signals transport in one direction deserves more exploration.
Although the nonreciprocal routers with only one output terminal have been proposed theoretically~\cite{LiGL2018PRA,DuL2020OE,Metelmann2018PRA,RenYl2022PRA} and realized experimentally in optomechanics~\cite{ShenZ2023PRL}, nonreciprocal routing with multiaccess channels has not been extensively studied yet.

In this paper, we introduce a nonreciprocal router with one transmitter, one receiver, and two output terminals, and propose a scheme to realized the nonreciprocal router based on an optomechanical plaquette composing of two optical modes and two mechanical modes.
We demonstrate that the optomechanical plaquette can serve as a nonreciprocal router with one optical mode as transmitter, another optical mode as receiver, and two mechanical modes as output terminals.
We will investigate the optimal conditions for nonreciprocal router and discuss its robustness against the imperfectness.
Our work may inspire the study of nonreciprocal quantum nodes for scalable quantum information processing in quantum secure network and communication.

The remainder of the paper is organized as follows. In Sec.~II, we introduces
the model of nonreciprocal router and show how to realize it in an optomechanical plaquette.
In Sec.~III, we show the nonreciprocal scattering matrix of the optomechanical plaquette and discuss
the influence of experimental parameters on the nonreciprocal routing effect.  
This work is summarized in Sec.~IV.

\section{Nonreciprocal router and optomechanical plaquette}

In this section, we will introduce the basic ideal of a nonreciprocal router, and then propose an optomechanical model to realize the nonreciprocal router.

\begin{figure}[tbp]
\includegraphics[bb=56 187 569 762, width=7 cm, clip]{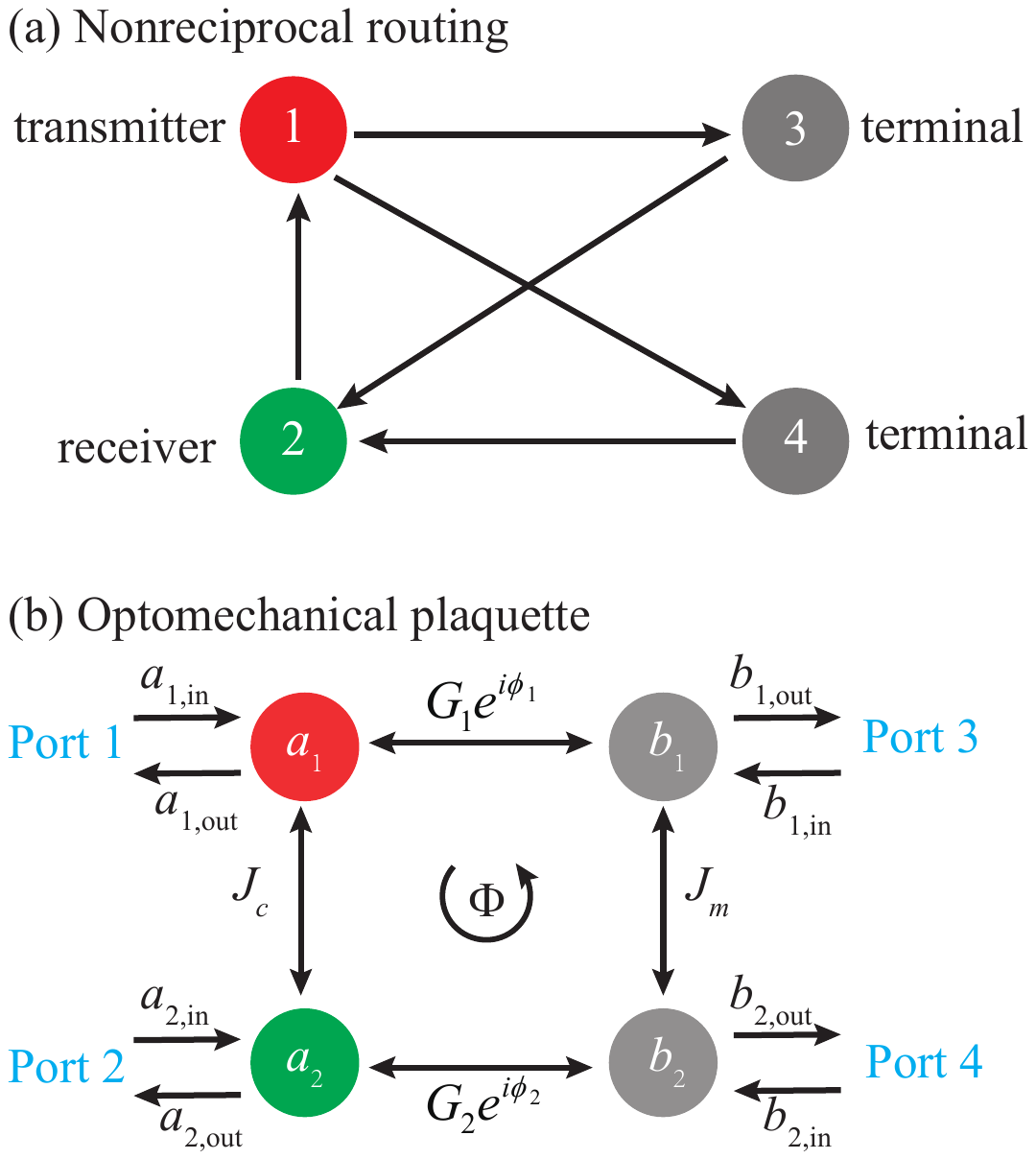}
\caption{(Color online) (a) Schematic diagram of a nonreciprocal router with one transmitter, one receiver, and two output terminals. The signals transport unidirectionally from the transmitter to the two terminals, from the terminals to the receiver, and from the receiver to the transmitter.
(b) Schematic diagram of a four-mode optomechanical plaquette comprising of two optical modes $(a_{1}$ and $a_{2})$
and two mechanical modes $(b_{1}$ and $b_{2})$.
The optical (mechanical) modes are coupled to each other with strength $J_{c}$ ($J_{m}$), and the optical and mechanical modes ($a_{1}$ and $b_{1}$, $a_{2}$ and $b_{2}$) are coupled
via radiation pressure, with a synthetic flux $\Phi=\phi_2 - \phi_1$ threading the four-mode plaquette for the linearized optomechancial coupling strengths $G_1 e^{i\phi_1}$ and $G_2 e^{i\phi_2}$. }
\label{fig1}
\end{figure}

\subsection{Nonreciprocal router} \label{NR}

We consider a nonreciprocal router consisting of one transmitter, one receiver, and two terminals. The basic ideal for the nonreciprocal router considered in the paper is shown in Fig.~\ref{fig1}(a).
Let's introduce the characteristics of the nonreciprocal router. The nonreciprocity indicates that the signals transport nonreciprocal between different ports, which are shown in the following three aspects: 
Firstly, the signal is transmitted from the transmitter to the terminals but cannot return from the terminals to the transmitter.
Secondly, the signal from the terminals can be transmitted to the receiver, but the signal from the receiver cannot transport to the terminals.
Thirdly, the nonreciprocity between the transmitter and the receiver is manifested as the signal can only be transmitted from the receiver to the transmitter but not from the transmitter to the receiver.
In order to realize such functions in an integrated platform, we consider an optomechanical plaquette, which has been used to realize non-reciprocal photon transport with
high isolation and directional optical amplification via synthetic magnetism~\cite{Fang2017NatPh,Mathew2020NatNa,Pino2022Natur,ChenY2021PRL,Peterson2017PRX,Bernier2017NatCo,Barzanjeh2017NatCo,Mercier2019PRAPP,Mercier2020PRL}.

\subsection{Optomechanical plaquette} \label{OMP}

We consider an optomechanical plaquette composing of two optical modes $a_{j}$ $\left(j=1,2\right) $ with frequency $\omega_{c,j}$ and two mechanical modes $b_{j}$ with frequency $\omega _{j}$, as shown in Fig.~\ref{fig1}(b). The two optical (mechanical) modes couple to each other with strength $J_{c}$ ($J_{m}$), and the optical mode ($a_{j}$) interacts with the mechanical mode ($b_{j}$) via radiation pressure with single-photon optomechanical coupling rate $g_{j}$. To enhance
the optomechanical coupling strength $g_{j}$, each optical mode is pumped by a strong optical field with strength $\varepsilon_{j}$ and frequency $\omega _{p,j}$ relatively detuned from the resonant frequency of optical mode $\omega_{c,j}$ by the mechanical frequency $\omega _{j}$, $\Delta_{0,j}\equiv \omega_{c,j}-\omega _{p,j} \approx \omega _{j}$.
In a rotating frame with respect to the unitary transformation $R(t)=\exp(-i\sum_{j=1,2} \omega _{p,j} t a_{j}^{\dag }a_{j})$, the optomechanical plaquette can be described by a Hamiltonian as $(\hbar =1)$%
\begin{eqnarray}
H &=&\sum_{j=1,2}\left[ \Delta _{0,j}a_{j}^{\dag }a_{j}+\omega
_{j}b_{j}^{\dag }b_{j}+g_{j}a_{j}^{\dag }a_{j}\left( b_{j}^{\dag
}+b_{j}\right) \right]  \nonumber \\
&&+J_{c}\left( a_{1}^{\dag }a_{2}+a_{1}a_{2}^{\dag }\right) +J_{m}\left(
b_{1}^{\dag }b_{2}+b_{1}b_{2}^{\dag }\right)  \nonumber \\
&&+i\left( \varepsilon _{1}a_{1}e^{i\varphi _{1}}+\varepsilon
_{2}a_{2}e^{i\varphi _{2}}-\mathrm{H.c.}\right) ,
\end{eqnarray}
where $\varphi _{1}$ and $\varphi _{2}$ are the phases of the pumping
fields and they are correlated, as an essential ingredient for nonreciprocity.

According to the Heisenberg equation of motion and taking into account the damping and
corresponding input noise, we get the quantum Langevin equations (QLEs) as
\begin{eqnarray}
\frac{d}{dt}a_{1} &=&\left\{ -\frac{\kappa }{2}-i\left[ \Delta
_{0,1}+g_{1}\left( b_{1}^{\dag }+b_{1}\right) \right] \right\} a_{1}
\nonumber \\
&&-iJ_{c}a_{2}-\varepsilon _{1}e^{-i\varphi _{1}}+\sqrt{\kappa _{e}}a_{1,%
\mathrm{in}}+\sqrt{\kappa _{0}}a_{1,\mathrm{vac}},
\end{eqnarray}%
\begin{eqnarray}
\frac{d}{dt}a_{2} &=&\left\{ -\frac{\kappa }{2}-i\left[ \Delta
_{0,2}+g_{2}\left( b_{2}^{\dag }+b_{2}\right) \right] \right\} a_{2}
\nonumber \\
&&-iJ_{c}a_{1}-\varepsilon _{2}e^{-i\varphi _{2}}+\sqrt{\kappa _{e}}a_{2,%
\mathrm{in}}+\sqrt{\kappa _{0}}a_{2,\mathrm{vac}},
\end{eqnarray}%
\begin{eqnarray}
\frac{d}{dt}b_{1} &=&\left( -\frac{\gamma }{2}-i\omega _{1}\right)
b_{1}-ig_{1}a_{1}^{\dag }a_{1}  \nonumber \\
&&-iJ_{m}b_{2}+\sqrt{\gamma _{e}}b_{1,\mathrm{in}}+\sqrt{\gamma _{0}}b_{1,%
\mathrm{th}},
\end{eqnarray}%
\begin{eqnarray}
\frac{d}{dt}b_{2} &=&\left( -\frac{\gamma }{2}-i\omega _{2}\right)
b_{2}-ig_{2}a_{2}^{\dag }a_{2}  \nonumber \\
&&-iJ_{m}b_{1}+\sqrt{\gamma _{e}}b_{2,\mathrm{in}}+\sqrt{\gamma _{0}}b_{2,%
\mathrm{th}},
\end{eqnarray}
where $\kappa_{e}$ and $\kappa_{0}$ ($\gamma_{e}$ and $\gamma_{0}$) are the external and internal decay rates of optical (mechanical) modes, with the corresponding input optical fields $a_{j, {\rm in}}$ and $a_{j,{\rm vac}}$ (mechanical fields $b_{j, {\rm in}}$ and $b_{j,{\rm th}}$).
$\kappa=\kappa _{e}+\kappa_{0}$ and $\gamma=\gamma _{e}+\gamma _{0}$ are the total decay rates of the optical and mechanical modes, respectively.

To obtain the linear response of the system to weak optical and mechanical signals, we will solve the QLEs by the standard process of the linearization~\cite{Aspelmeyer2014RMP}.
We solve the QLEs in steady state based on the mean field approximation first, then the linearization of the system near the steady-state values yields to the corresponding linearized QLEs.
The steady-state values $\alpha_{j}\equiv \langle a_j \rangle$ and $\beta _{j}\equiv \langle b_j \rangle$ are obtained from the QLEs by setting the time derivative terms to zero and using the factorization assumption $\langle A B \rangle=\langle A\rangle \langle B \rangle$ for any two operators $A$ and $B$, as
\begin{equation}
\alpha _{1}=e^{-i\varphi _{1}}\frac{\left( -\frac{\kappa }{2}-i\Delta
_{2}\right) \varepsilon _{1}+iJ_{c}\varepsilon _{2}e^{-i\left( \varphi
_{2}-\varphi _{1}\right) }}{\left( -\frac{\kappa }{2}-i\Delta _{2}\right)
\left( -\frac{\kappa }{2}-i\Delta _{1}\right) +J_{c}^{2}},
\end{equation}%
\begin{equation}
\alpha _{2}=e^{-i\varphi _{2}}\frac{\left( -\frac{\kappa }{2}-i\Delta
_{1}\right) \varepsilon _{2}+iJ_{c}\varepsilon _{1}e^{i\left( \varphi
_{2}-\varphi _{1}\right) }}{\left( -\frac{\kappa }{2}-i\Delta _{2}\right)
\left( -\frac{\kappa }{2}-i\Delta _{1}\right) +J_{c}^{2}},
\end{equation}%
where $\Delta _{j}\equiv \Delta _{0,j}+g_{j}\left( \beta _{j}^{\ast }+\beta _{j}\right) $.
Suppose $\varepsilon _{1}\sim \varepsilon _{2}$ and $\left\vert
\Delta _{j}\right\vert \gg J_{c}$, then we have%
\begin{equation}\label{SF}
\alpha _{j}= \vert \alpha _{j} \vert e^{i\phi_j} \approx \frac{\varepsilon _{j}e^{-i\varphi _{j}}}{\left( -\frac{%
\kappa }{2}-i\Delta _{j}\right) },
\end{equation}%
which means that the amplitude $\vert \alpha _{j} \vert$ and phase $\phi_j$ of $\alpha _{j}$ can be freely adjusted by the external field $\varepsilon _{j}e^{-i\varphi _{j}}$.

The linearized QLEs for the quantum fluctuation operators $\delta a_{j}\equiv a_{j}-\alpha _{j}$ and $\delta b_{j}\equiv b_{j}-\beta _{j}$ $(j=1,2)$
are obtained as
\begin{eqnarray}
\frac{d}{dt}\delta a_{1} &=&\left( -\frac{\kappa }{2}-i\Delta _{1}\right)
\delta a_{1}-iJ_{c}\delta a_{2}-iG_{1}e^{i\phi _{1}}\delta b_{1}  \nonumber
\\
&&+\sqrt{\kappa _{e}}a_{1,\mathrm{in}}+\sqrt{\kappa _{0}}a_{1,\mathrm{vac}},
\end{eqnarray}%
\begin{eqnarray}
\frac{d}{dt}\delta a_{2} &=&\left( -\frac{\kappa }{2}-i\Delta _{2}\right)
\delta a_{2}-iJ_{c}\delta a_{1}-iG_{2}e^{i\phi _{2}}\delta b_{2}  \nonumber
\\
&&+\sqrt{\kappa _{e}}a_{2,\mathrm{in}}+\sqrt{\kappa _{0}}a_{L,\mathrm{vac}},
\end{eqnarray}%
\begin{eqnarray}
\frac{d}{dt}\delta b_{1} &=&\left( -\frac{\gamma }{2}-i\omega _{1}\right)
\delta b_{1}-iJ_{m}\delta b_{2}-iG_{1}e^{-i\phi _{1}}\delta a_{1}  \nonumber
\\
&&+\sqrt{\gamma _{e}}b_{1,\mathrm{in}}+\sqrt{\gamma _{0}}b_{1,\mathrm{th}},
\end{eqnarray}%
\begin{eqnarray}
\frac{d}{dt}\delta b_{2} &=&\left( -\frac{\gamma }{2}-i\omega _{2}\right)
\delta b_{2}-iJ_{m}\delta b_{1}-iG_{2}e^{-i\phi _{2}}\delta a_{2}  \nonumber
\\
&&+\sqrt{\gamma _{e}}b_{2,\mathrm{in}}+\sqrt{\gamma _{0}}b_{2,\mathrm{th}}.
\end{eqnarray}%
Here $G_{1}=\left\vert \alpha _{1}\right\vert g_{1}$ and $G_{2}=\left\vert \alpha _{2}\right\vert g_{2}$ are the linearized optomechancial
coupling strengths; the nonlinear terms are negligible for the assumption $\vert \alpha _{j}\vert^2 \gg \langle \delta a^{\dag }_j\delta a_j \rangle$, and the counter-rotating terms are neglected based on rotation-wave approximation under conditions $\Delta _{j} \approx \omega_{j} \gg G_{j}$.

The linearized QLEs can be concisely expressed as%
\begin{equation}\label{LQLEs}
\frac{d}{dt}V\left( t\right) =-MV\left( t\right) +\sqrt{\Gamma _{e}}V_{%
\mathrm{in}}\left( t\right) +\sqrt{\Gamma _{0}}V_{\mathrm{noise}}\left(
t\right) ,
\end{equation}%
where $V\left( t\right) $ is the vector for the quantum fluctuation
operators defined by $V\left( t\right) \equiv \left( \delta a_{1},\delta
a_{2},\delta b_{1},\delta b_{2}\right) ^{T}$, $V_{\mathrm{in}}\left(
t\right) \equiv \left( a_{1,\mathrm{in}},a_{2,\mathrm{in}},b_{1,\mathrm{in}},b_{2,%
\mathrm{in}}\right) ^{T}$, $V_{\mathrm{noise}}\left( t\right) \equiv \left(
a_{1,\mathrm{vac}},a_{2,\mathrm{vac}},b_{1,\mathrm{th}},b_{2,\mathrm{th}%
}\right) ^{T}$, $\Gamma _{e} \equiv \mathrm{diag}(\kappa _{e},\kappa _{e},\gamma
_{e},\gamma _{e})$, $\Gamma _{0} \equiv \mathrm{diag}(\kappa _{0},\kappa
_{0},\gamma _{0},\gamma _{0})$, and the linearized coefficient matrix $M$ is
given by%
\begin{equation}
M=\left(
\begin{array}{cccc}
\frac{\kappa }{2}+i\Delta _{1} & iJ_{c} & iG_{1}e^{i\phi _{1}} & 0 \\
iJ_{c} & \frac{\kappa }{2}+i\Delta _{2} & 0 & iG_{2}e^{i\phi _{2}} \\
iG_{1}e^{-i\phi _{1}} & 0 & \frac{\gamma }{2}+i\omega _{1} & iJ_{m} \\
0 & iG_{2}e^{-i\phi _{2}} & iJ_{m} & \frac{\gamma }{2}+i\omega _{2}%
\end{array}%
\right) .
\end{equation}
In order to ensure the stability of the system, we need to ensure that the real part of all eigenvalues of the coefficient matrix $M$ are positive in the following discussions.

The linearized QLEs can be solved analytically in the frequency domain by the method of Fourier transform, with the definition of Fourier transform for operator $o$ as
\begin{equation}
\widetilde{o}\left( \omega \right) =\frac{1}{\sqrt{2\pi }}\int_{-\infty
}^{+\infty }o\left( t\right) e^{i\omega t}dt.
\end{equation}%
The solution of the linearized QLEs in the frequency domain is obtained as%
\begin{equation}
\widetilde{V}\left( \omega \right) =\left( M-i\omega I\right) ^{-1}\left[
\sqrt{\Gamma _{e}}\widetilde{V}_{\mathrm{in}}\left( \omega \right) +\sqrt{%
\Gamma _{0}}\widetilde{V}_{\mathrm{noise}}\left( \omega \right) \right] ,
\end{equation}%
where $I$ denotes the identity matrix. Based on the input-output relation~\cite{Gardiner1985PRA}
\begin{equation}
\widetilde{a}_{j,\mathrm{out}}\left( \omega \right) +\widetilde{a}_{j,%
\mathrm{in}}\left( \omega \right) =\sqrt{\kappa _{e}}\delta \widetilde{a}%
_{j}\left( \omega \right) ,
\end{equation}%
\begin{equation}
\widetilde{b}_{j,\mathrm{out}}\left( \omega \right) +\widetilde{b}_{j,%
\mathrm{in}}\left( \omega \right) =\sqrt{\gamma _{e}}\delta \widetilde{b}%
_{j}\left( \omega \right) ,
\end{equation}%
the output fields from the optomechanical system are obtained as
\begin{equation}
\widetilde{V}_{\mathrm{out}}\left( \omega \right) =U\left( \omega \right)
\widetilde{V}_{\mathrm{in}}\left( \omega \right) +W \left( \omega
\right) \widetilde{V}_{\mathrm{noise}}\left( \omega \right) ,
\end{equation}%
where
\begin{equation}\label{UU}
U\left( \omega \right) =\sqrt{\Gamma _{e}}\left( M-i\omega I\right) ^{-1}%
\sqrt{\Gamma _{e}}-I,
\end{equation}%
\begin{equation}
W\left( \omega \right) =\sqrt{\Gamma _{e}}\left( M-i\omega
I\right) ^{-1}\sqrt{\Gamma _{0}}
\end{equation}%
are the scattering matrices.
So the scattering probabilities from Port $j$ to Port $i$ are given by%
\begin{equation}
S_{ij}\left( \omega \right) =\left\vert U_{ij}\left( \omega \right)
\right\vert ^{2},
\end{equation}
where $U_{ij}\left( \omega \right) $ $\left( i,j=1,2,3,4\right)$ denotes
the element of $U\left( \omega \right)$ at $i$th row and $j$th column, given analytically in Appendix.

\subsection{Parameter conditions for nonreciprocal routing}\label{OP}

The optomechanical plaquette can work as a nonreciprocal router with two optical modes acting as the transmitter and receiver, and two mechanical modes as two terminals.
For example, we can use the optical mode $a_1$ as the transmitter, $a_2$ as the receiver, and the two mechanical modes $b_1$ and $b_2$ as two terminals.
In order to realize the nonreciprocal router shown in Fig.~\ref{fig1}(a), the scattering probabilities from Port $1$ to $2$, from Ports $3$ and $4$ to $1$, and from Port $2$ to $3$ and $4$, should be significantly inhibited, i.e., approximately equal to zero,
\begin{eqnarray}\label{CD1}
&& S_{21}\left( \omega \right)\approx 0, \nonumber \\
&& S_{13}\left( \omega \right)\approx S_{14}\left( \omega \right)\approx 0, \\
&& S_{32}\left( \omega \right)\approx S_{42}\left( \omega \right)\approx 0. \nonumber
\end{eqnarray}%
Besides that, we can also use the optical mode $a_2$ as the transmitter, $a_1$ as the receiver, and the two mechanical modes as two terminals.
Then the scattering probabilities from Port $2$ to $1$, from Ports $3$ and $4$ to $2$, and from Port $1$ to $3$ and $4$, should be significantly inhibited as
\begin{eqnarray}\label{CD2}
&& S_{12}\left( \omega \right)\approx 0, \nonumber \\
&& S_{31}\left( \omega \right)\approx S_{41}\left( \omega \right)\approx 0, \\
&& S_{23}\left( \omega \right)\approx S_{24}\left( \omega \right)\approx 0. \nonumber
\end{eqnarray}%
The parameter conditions for these scattering probabilities given in Eqs.~(\ref{CD1}) and (\ref{CD2}) can be derived from the analytical expression of $U_{ij}\left( \omega \right) $ given in Appendix.

\begin{table}[tbp]
\centering
\caption{The inhibited scattering paths for different synthetic flux $\Phi$ and resonant frequency $\omega$. The other parameters satisfy the conditions $\omega_m \gg \kappa=2J_c\gg J_m\gg \gamma$, $G^2= J_c \gamma \ll \kappa J_m$, where $\omega_1=\omega_2=\omega_m$, $\Delta_1=\Delta_2=\omega_m$, and $G_1=G_2=G$.} \label{Tab1}
\begin{tabular}{c|c|c}
 \hline
&  $\omega =\omega_m - J_m$    &  $\omega =\omega_m+ J_m$  \\ \hline

\multirow{3}{*}{$\Phi = \pi/2$}           &$S_{21}(\omega)\approx 0$     & $S_{12}(\omega)\approx 0$  \\

&$S_{13}(\omega)\approx S_{14}(\omega)\approx 0$     & $S_{31}(\omega)\approx S_{41}(\omega)\approx 0$  \\

&$S_{32}(\omega)\approx S_{42}(\omega)\approx 0$     & $S_{23}(\omega)\approx S_{24}(\omega)\approx 0$  \\  \hline

\multirow{3}{*}{$\Phi = 3\pi/2$}             &$S_{12}(\omega)\approx 0$     & $S_{21}(\omega)\approx 0$  \\

&$S_{31}(\omega)\approx S_{41}(\omega)\approx 0$     & $S_{13}(\omega)\approx S_{14}(\omega)\approx 0$  \\

&$S_{23}(\omega)\approx S_{24}(\omega)\approx 0$     & $S_{32}(\omega)\approx S_{42}(\omega)\approx 0$  \\  \hline

\end{tabular}
\end{table}

Before the numerical simulation given in the next section, let us derive the parameters conditions for nonreciprocal router analytically based on some assumptions.
Without loss of generality, here we set $\omega_1=\omega_2=\omega_m$, $\Delta_1=\Delta_2=\omega_m$, and $G_1=G_2=G$ for simplicity.
Moreover, we introduce the synthetic flux (relative phase) $\Phi \equiv \phi _{2}-\phi _{1}$ threading the four-mode plaquette to break time-reversal symmetry of the system, leading to nonreciprocal transport of signals between different ports.
Based on the Eqs.~(\ref{A1})-(\ref{A16}), we find that the optomechanical plaquette can work as a nonreciprocal router with $a_1$ as the transmitter and $a_2$ as the receiver around the frequency $\omega=\omega_m-J_m$ for $\Phi=\pi/2$ or around the frequency $\omega=\omega_m+J_m$ for $\Phi=3\pi/2$.
Conversely, the optomechanical plaquette can also work as a nonreciprocal router with $a_1$ as the receiver and $a_2$ as the transmitter around the frequency $\omega=\omega_m+J_m$ for $\Phi=\pi/2$ or around the frequency $\omega=\omega_m-J_m$ for $\Phi=3\pi/2$.
The other parameter conditions are given by: \\
(i) to realize $S_{12}\left( \omega \right)\approx 0$ or $S_{21}\left( \omega \right)\approx 0$, the parameters need to satisfy the conditions 
\begin{equation}
J_m \gg \gamma, \qquad G_1=G_2=\sqrt{J_c \gamma};
\end{equation}
(ii) $S_{14}\left( \omega \right)\approx S_{32}\left( \omega \right)\approx 0$ or $S_{41}\left( \omega \right)\approx S_{23}\left( \omega \right)\approx 0$, are expected with the parameters satisfying the conditions 
\begin{equation}
\kappa \gg J_m \gg \gamma, \qquad \kappa=2 J_c;
\end{equation}
(iii) $S_{13}\left( \omega \right)\approx S_{42}\left( \omega \right)\approx 0$ or $S_{31}\left( \omega \right)\approx S_{24}\left( \omega \right)\approx 0$, are obtained under the conditions 
\begin{equation}
\kappa \gg J_m \gg \gamma, \quad G^2\ll \kappa J_m,\quad \kappa=2 J_c.
\end{equation}
The conditions for nonreciprocal routing are summarized in Table~\ref{Tab1}.

The conditions required to observe nonreciprocal routing can be reached for the current state-of-the-art optomechanical crystal circuit systems.
For example, Ref.~\cite{Fang2017NatPh} demonstrated nonreciprocal photon transport and amplification by an optomechanical plaquette with the parameters:
mechanical frequency $\omega_m/2\pi \approx 5.8$ GHz, external optical damping rate $\kappa_e/2\pi \approx 0.75 \sim 1$ GHz, intrinsic optical damping rate $\kappa_0/2\pi \approx 0.3$ GHz,
single-photon optomechanical coupling rate $g_i/2 \pi \approx 0.8$ MHz, mechanical dissipation rate $\gamma/2\pi \approx 4 \sim 5$ MHz, optical hopping rate $J_c/2\pi \approx 0.1 \sim 1.4$ GHz,
mechancial hopping rate $J_m/2\pi \approx 2.8$ MHz. The coupling rates $J_c$ and $J_m$ can be varied by changing the number and shape of the holes that make up the optomechanical crystal between the two optomechanical cavities~\cite{Burgwal2023NatCo}. The linearized optomechanical coupling rates $G_j=\left\vert \alpha _{j}\right\vert g_j$ can be enhanced by pumping the optomechanical cavities with phase-correlated lasers.

\section{Nonreciprocal Routing}

In this section, we will show the nonreciprocal routing effect in the optomechanical plaquette by numerical calculations.
As a simple example, we show the nonreciprocal scattering matrix first. Then, we discuss the optimal parameters (synthetic flux $\Phi$, optomechanical coupling rate $G$, and external decay rate $\kappa_e$) for nonreciprocal routing, and the influences of the detuning between the two mechanical modes ($\delta=\omega_1-\omega_2$) and internal optical and mechanical decays ($\kappa_0$ and $\gamma_0$) on nonreciprocal routing.

\subsection{Nonreciprocal scattering matrix}

\begin{figure}[tbp]
\includegraphics[bb=27 10 570 814, width=8.5 cm, clip]{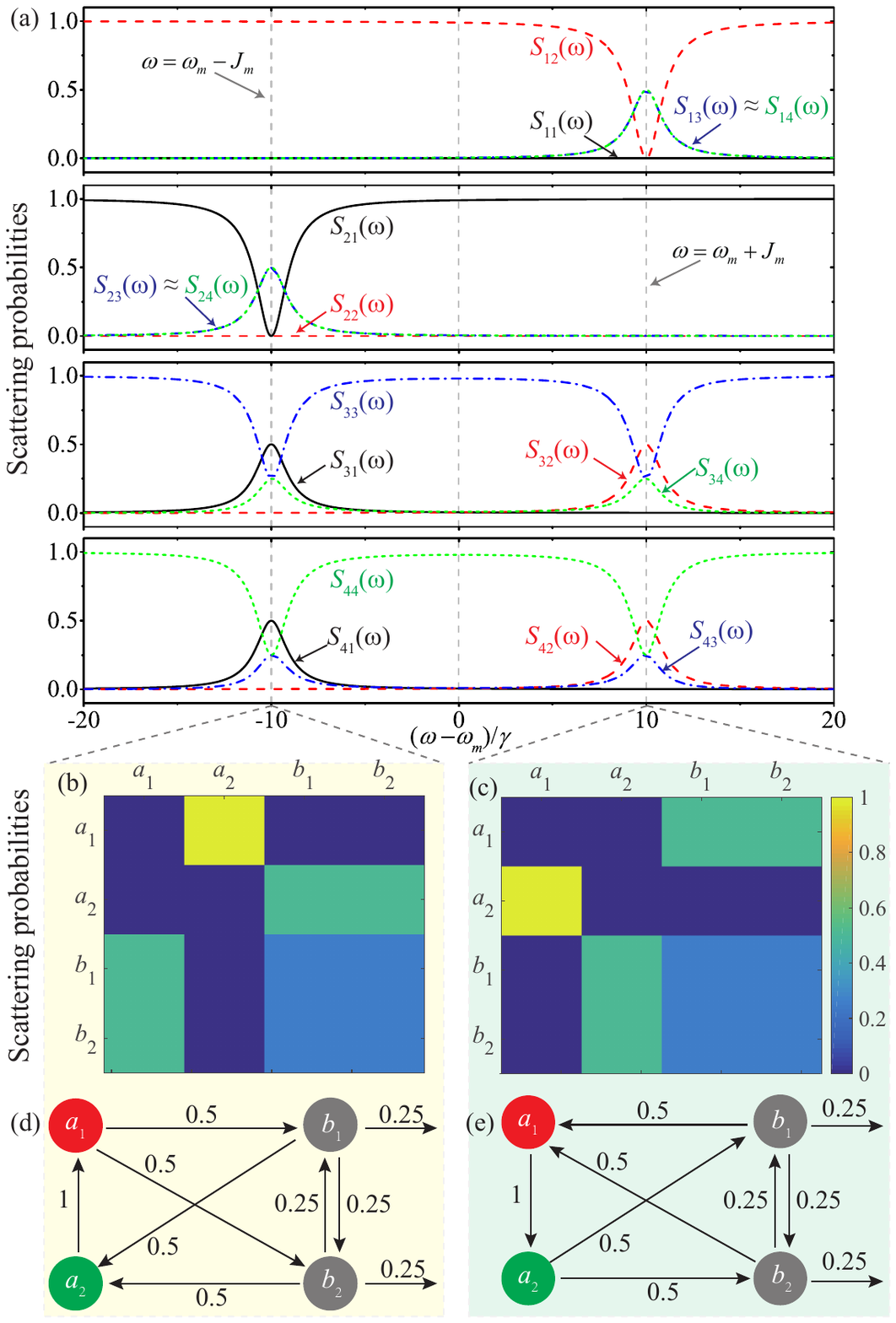}
\caption{(Color online) Scattering probabilities between different ports. (a) $S_{ij}(\omega)$ verses frequency $(\omega-\omega_m)/\gamma$. Scattering matrices for (b) $\omega=\omega_m-J_m$ and (c) $\omega=\omega_m+J_m$, with the corresponding probability flow charts given in (d) and (e), respectively. The other parameters are $\omega_{1}=\omega_{2}=\omega_{m}$, $\Delta_{1}=\Delta_{2}=\omega_{m}$, $J_m=10 \gamma$, $J_c=500\gamma$, $\gamma_{0}=\kappa_{0}=0$, $\gamma_{e}=\gamma$, $\kappa_{e}=\kappa$, $\Phi=\pi/2$, $G_1=G_2=\sqrt{J_c \gamma}$, and $\kappa=2J_c$. }
\label{fig2}
\end{figure}

The scattering probabilities $S_{ij}(\omega)$ between different ports are shown as functions of the frequency $(\omega-\omega_m)/\gamma$ in Fig.~\ref{fig2}(a).
The scattering probabilities $S_{ij}(\omega)$ change only round the splitting frequencies $\omega_m \pm J_m$ for the two mechanical modes $\omega_1=\omega_2=\omega_m$ with resonant interaction rate $J_m$.
Specifically, at frequency $\omega=\omega_m-J_m$, the signals can transport from optical mode $a_2$ to $a_1$ with nearly one hundred percent probability, i.e., $S_{12}(\omega)\approx 1$;
there is a 50/50 chance for the signals transport from $a_1$ either to $b_1$ or to $b_2$, i.e., $S_{31}(\omega)\approx S_{41}(\omega)\approx 0.5$;
the probability for the signals transport from $b_1$ or $b_2$ to $a_2$ is both about 50 percent, i.e., $S_{23}(\omega)\approx S_{24}(\omega)\approx 0.5$; all the transmission processes mentioned here are unidirectional, i.e., $S_{21}(\omega)\approx S_{14}(\omega)\approx S_{13}(\omega)\approx S_{42}(\omega)\approx S_{32}(\omega)\approx 0$.
In other words, the optomechanical plaquette works as nonreciprocal router, with optical mode $a_1$ as transmitter, $a_2$ as receiver, and the two mechanical modes $b_1$ and $b_2$ as two terminals.
In order to shown the nonreciprocal routing behavior even more clearly, the scattering matrix for $\omega=\omega_m-J_m$ are shown in Fig.~\ref{fig2}(b) and the corresponding probability flow charts given in (d).
It is obvious that the scattering matrix is asymmetry, and the probability flow are unidirectional.

In contrast, at frequency $\omega=\omega_m+J_m$, the optomechanical plaquette exhibits nonreciprocal routing behavior with signals transporting in different direction.
As shown in Figs.~\ref{fig2}(c) and \ref{fig2}(e), the optical mode $a_2$ is taken as a transmitter and the optical mode $a_1$ is taken as a receiver.
As summarized in Table~\ref{Tab1}, the optomechanical plaquette show nonreciprocal routing behavior in revers direction for $\Phi=\pi/2$ and $\Phi=3\pi/2$. Thus we can steer the direction of the nonreciprocal routing by adjusting either the synthetic flux $\Phi$ or the frequency of the input signals.

\subsection{Numerical results for optimal parameters}

\begin{figure*}[htbp]
\includegraphics[bb=19 273 539 658, width=16 cm, clip]{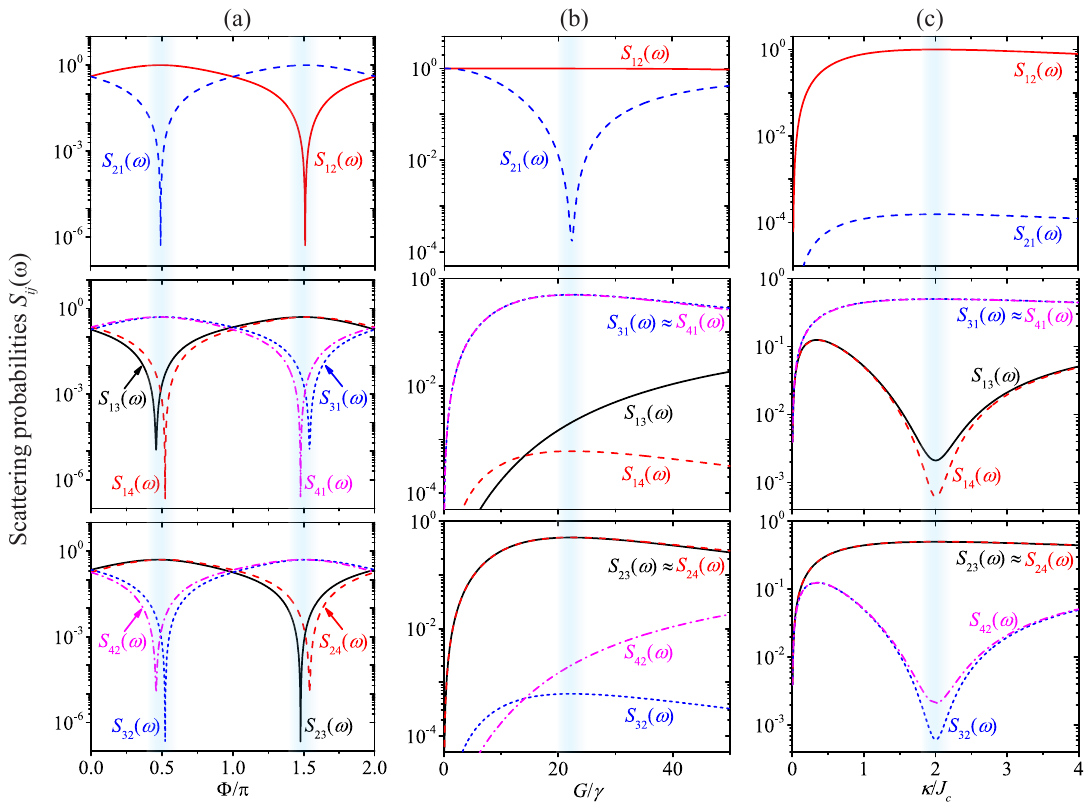}
\caption{(Color online) The optimal parameters for nonreciprocal routing. Scattering probabilities $S_{ij}(\omega)$ are plotted as functions of (a) the synthetic flux $\Phi/\pi$ for $\kappa=2J_c$ and $G_1=G_2=\sqrt{J_c \gamma}$, (b) the optomechancial coupling strength $G_1=G_2\equiv G$ for $\Phi=\pi/2$ and $\kappa=2J_c$, and (c) the optical decay rate $\kappa/J_c$ for $\Phi=\pi/2$ and $G_1=G_2=\sqrt{J_c \gamma}$.
The parameters are $\omega_{1}=\omega_{2}=\omega_{m}$, $\Delta_{1}=\Delta_{2}=\omega_{m}$, $J_m=10 \gamma$, $J_c=500\gamma$, $\omega=\omega_m-J_m$, $\gamma_{0}=\kappa_{0}=0$, $\gamma_{e}=\gamma$, and $\kappa_{e}=\kappa$.}
\label{fig3}
\end{figure*}

In order to demonstrate the optimal conditions for nonreciprocal routing quantitatively, i.e., $\kappa=2 J_c$, $G=\sqrt{J_c \gamma}$, and $\Phi=\pi/2$ or $3\pi/2$, we show the scattering probabilities $S_{ij}(\omega)$ as functions of synthetic flux $\Phi$, linearized optomechancial
coupling strength $G$, and external optical decay $\kappa_e$ in Figs.~\ref{fig3}(a)-\ref{fig3}(c) under the conditions $\omega_m \gg \kappa\gg J_m\gg \gamma$.

Synthetic flux $\Phi$ is one ingredient for breaking the time-reversal symmetry, and plays a key role in controlling the nonreciprocal routing [see Fig.~\ref{fig3}(a)]. Agree with the analytically results given in Table~\ref{Tab1} at frequency $\omega=\omega_m-J_m$, we have $S_{12}(\omega)\approx 1 \gg S_{21}(\omega)$ around $\Phi=\pi/2$ and $S_{12}(\omega)\ll S_{21}(\omega)\approx 1$ around $\Phi=3\pi/2$.
Similarly, we have $S_{31}(\omega)\approx S_{41}(\omega)\approx S_{23}(\omega)\approx S_{24}(\omega)\approx 0.5 \gg \{S_{13}(\omega), S_{14}(\omega), S_{32}(\omega), S_{42}(\omega)\}$ around $\Phi=\pi/2$ and $S_{13}(\omega)\approx S_{14}(\omega)\approx S_{32}(\omega)\approx S_{42}(\omega)\approx 0.5 \gg \{S_{31}(\omega), S_{41}(\omega), S_{23}(\omega), S_{24}(\omega)\}$ around $\Phi=3\pi/2$.
So we can steer the transmission direction of the signals in the optomechanical plaquette by tuning the synthetic flux $\Phi$, which can be freely adjusted by the external pumping lasers given in Eq.~(\ref{SF}).

The scattering probabilities between Port $1$ and $2$ [$S_{12}(\omega)/S_{21}(\omega)$] show different behaviors with the ones between other Ports [$S_{31}(\omega)/S_{13}(\omega)$, $S_{41}(\omega)/S_{14}(\omega)$, $S_{23}(\omega)/S_{32}(\omega)$, and $S_{24}(\omega)/S_{42}(\omega)$].
According to the analytical results given in Sec.~\ref{OP}, in order to realize $S_{12}(\omega)\approx 1 \gg S_{21}(\omega)$, one required condition is $G_1=G_2=G=\sqrt{J_c \gamma}$.
This condition agrees well with the numerical results shown in Fig.~\ref{fig3}(b) (up). There is a dip in $S_{21}(\omega)$ around the value of $G\approx 22.4 \gamma \approx \sqrt{J_c \gamma}$, with $S_{12}(\omega) \approx 1$ for a wide range of $G$.
In contrast, the isolations [$S_{31}(\omega)/S_{13}(\omega)$, $S_{41}(\omega)/S_{14}(\omega)$, $S_{23}(\omega)/S_{32}(\omega)$, and $S_{24}(\omega)/S_{42}(\omega)$] decrease with the increasing of $G$. Nevertheless, we can obtain the maximal values of $S_{31}(\omega)\approx S_{41}(\omega)\approx S_{23}(\omega) \approx S_{24}(\omega) \approx 0.5$ under the condition $G\approx \sqrt{J_c \gamma}$ [see Fig.~\ref{fig3}(b) (middle and down)].

As shown in Fig.~\ref{fig3}(c) (middle and down), there is a dip for $\{S_{13}(\omega), S_{14}(\omega), S_{32}(\omega), S_{42}(\omega)\}$ with $S_{31}(\omega)$, $S_{41}(\omega)$, $S_{23}(\omega)$, and $S_{24}(\omega)$ reaching the maximal value $0.5$ around the point $\kappa \approx 2 J_c$.
These agree well with the required condition obtained analytically in Sec.~\ref{OP}.
Differently, both $S_{12}(\omega)$ and $S_{21}(\omega)$ increase with the creasing of $\kappa/J_c$ and reach the maximal value around the point $\kappa \approx 2 J_c$, and the isolation $S_{12}(\omega)/S_{21}(\omega)$ is not sensitive to the change of $\kappa/J_c$ [see Fig.~\ref{fig3}(c) (up)].

\subsection{Effects of detuning and internal decays}

\begin{figure*}[tbp]
\includegraphics[bb=19 273 544 658, width=16 cm, clip]{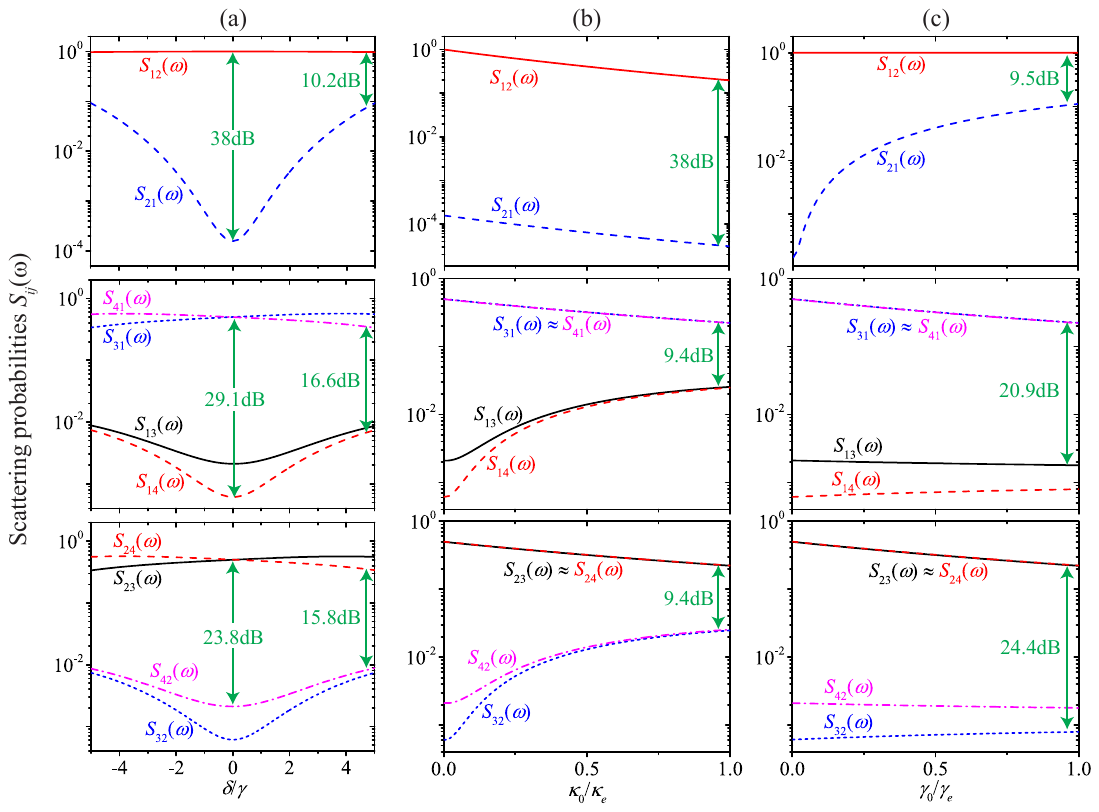}
\caption{(Color online) Effects of frequency difference and intrinsic decay on the nonreciprocal routing. Scattering probabilities $S_{ij}(\omega)$ are plotted as functions of (a) the frequency difference of two mechanical modes $\delta=\omega_1-\omega_2$ for $\kappa_0=\gamma_0=0$, (b) the intrinsic optical decay $\kappa_0/\kappa_e$ for $\delta =\gamma_0=0$, and (c) the intrinsic mechanical decay $\gamma_0/\gamma_e$ for $\delta=\kappa_0=0$.
The parameters are $\omega_{m}=(\omega_{1}+\omega_{2})/2$, $\Delta_{1}=\Delta_{2}=\omega_{m}$, $\Phi=\pi/2$, $J_m=10 \gamma_e$, $J_c=500\gamma_e$, $\kappa_e=2J_c$, $G_1=G_2=\sqrt{J_c\gamma_e}$, and $\omega=\omega_m-J_m$.}
\label{fig4}
\end{figure*}

In the discussions above, we assume that the two mechanical modes have the same resonant frequency $\omega_{1}=\omega_{2}=\omega_{m}$ and ignore the intrinsic decay $\kappa_0=\gamma_0=0$ for simplicity.
However, there are unpredictable uncertainty in devices fabrication and we may have the frequency difference of two mechanical modes $\delta=\omega_1-\omega_2$ and intrinsic decay $\kappa_0$ and $\gamma_0$ in experiments.
In this subsection, we will show numerically that the nonreciprocal routing is robust against the mechanical frequency difference $\delta$ and intrinsic decay $\kappa_0$ and $\gamma_0$.

The scattering probabilities are plotted as functions of the mechanical frequency difference $\delta/\gamma$ in Fig.~\ref{fig4}(a).
$S_{21}(\omega)$ is sensitive to the mechanical frequency difference $\delta/\gamma$, but $S_{12}(\omega)$ is almost unchanged.
The isolation of $S_{12}(\omega)/S_{21}(\omega)$ decrease from $38$dB to $10.2$dB with the increasing of $\delta$ from $0$ to $\pm 5\gamma$.
The isolations of $S_{41}(\omega)/S_{14}(\omega)$, $S_{31}(\omega)/S_{13}(\omega)$, $S_{23}(\omega)/S_{32}(\omega)$, and $S_{24}(\omega)/S_{42}(\omega)$ are about $16$dB at the mechanical frequency difference $\delta=\pm 5\gamma$.

In order to study the influences of the intrinsic optical decay on the nonreciprocal routing, the scattering probabilities are plotted as functions of $\kappa_0/\kappa_e$ in Fig.~\ref{fig4}(b).
Both $S_{12}(\omega)$ and $S_{21}(\omega)$ decrease with the creasing of $\kappa_0$, but the isolation of $S_{12}(\omega)/S_{21}(\omega)$ is not sensitive to the change of $\kappa_0$ [see Fig.~\ref{fig4}(b) (up)].
The isolations of $S_{41}(\omega)/S_{14}(\omega)$, $S_{31}(\omega)/S_{13}(\omega)$, $S_{23}(\omega)/S_{32}(\omega)$, and $S_{24}(\omega)/S_{42}(\omega)$ decrease with the increasing of $\kappa_0$, and reach about $9.4$dB with intrinsic decay $\kappa_0=\kappa_e$.

The influences of the intrinsic mechanical decay on the nonreciprocal routing is shown in Fig.~\ref{fig4}(c).
The scattering probability $S_{21}(\omega)$ increases with the creasing of $\gamma_0$, and the isolation of $S_{12}(\omega)/S_{21}(\omega)$ decreases from $38$dB to $9.5$dB with the increasing of $\gamma_0$ from $0$ to $\gamma_e$.
In contrast, $S_{31}(\omega)\approx S_{41}(\omega)$ and $S_{23}(\omega)\approx S_{24}(\omega)$ decease slowly with the increasing of $\gamma_0$, and $S_{13}(\omega)$, $S_{14}(\omega)$, $S_{42}(\omega)$, and $S_{32}(\omega)$ are not sensitive to $\gamma_0$.

\subsection{Nonreciprocal routing with normal mechanical modes}

\begin{figure*}[tbp]
\includegraphics[bb=41 326 532 628, width=18 cm, clip]{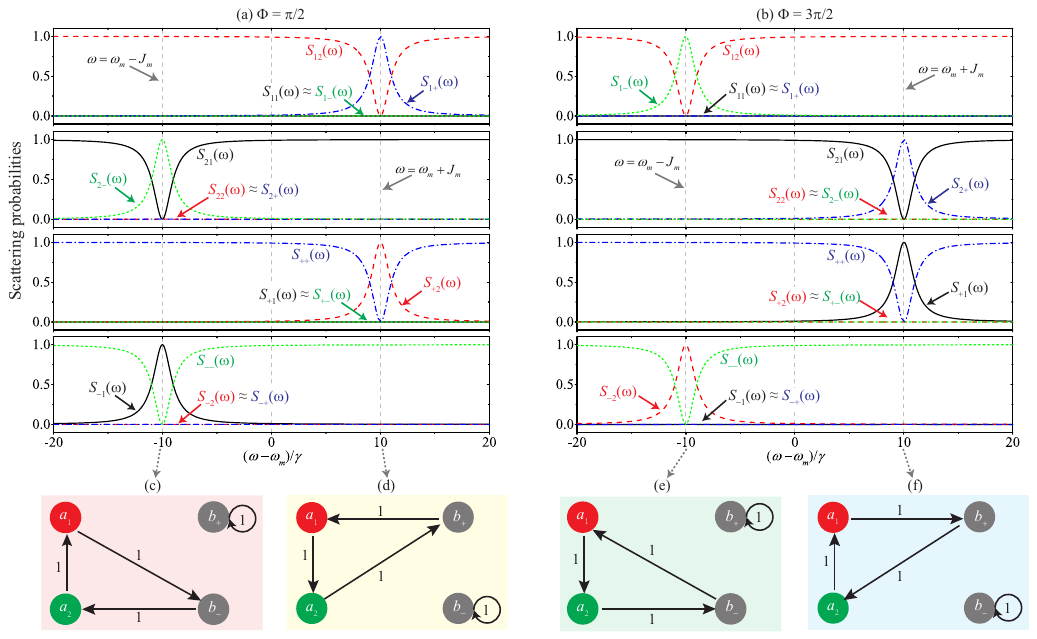}
\caption{(Color online) Scattering probabilities between different modes $S_{ij}(\omega)$ ($i,j=1,2,+,-$) verses frequency $(\omega-\omega_m)/\gamma$ for (a) $\Phi=\pi/2$ and (b) $\Phi=3\pi/2$, with the corresponding probability flow charts shown in (c)-(f). The parameters are $\omega_{1}=\omega_{2}=\omega_{m}$, $\Delta_{1}=\Delta_{2}=\omega_{m}$, $J_m=10 \gamma$, $J_c=500\gamma$, $\gamma_{0}=\kappa_{0}=0$, $\gamma_{e}=\gamma$, $\kappa_{e}=\kappa$, $G_1=G_2=\sqrt{J_c \gamma}$, and $\kappa=2J_c$.}
\label{fig5}
\end{figure*}

Under the strong coupling condition $J_m\gg \gamma$, we can introduce the normal modes ($b_{\pm}$) of the two coupled mechanical modes ($b_{1}$ and $b_{2}$) as
\begin{equation}
\delta b_{\pm }=\frac{1}{\sqrt{2}}\left( \delta b_{1}\pm \delta b_{2}\right),
\end{equation}%
with resonant frequency $\omega_\pm =\omega_m \pm J_m$.
The resonant frequency of $\omega_\pm$ provide us a clue to understand the scattering probabilities $S_{ij}(\omega)$ change dramatically round the splitting frequencies $\omega_m \pm J_m$.

The linearized QLEs~(\ref{LQLEs}) for the system can be rewritten with normal mechanical modes ($b_{\pm}$) as 
\begin{equation}
\frac{d}{dt}V^{\prime }\left( t\right) =-M^{\prime }V^{\prime }\left(
t\right) +\sqrt{\Gamma _{e}}V_{\mathrm{in}}^{\prime }\left( t\right) +\sqrt{%
\Gamma _{0}}V_{\mathrm{noise}}^{\prime }\left( t\right) ,
\end{equation}%
with $V^{\prime }\left( t\right) \equiv \left( \delta a_{1},\delta
a_{2},\delta b_{+},\delta b_{-}\right) ^{T}$, $V_{\mathrm{in}}^{\prime
}\left( t\right) =\left( a_{1,\mathrm{in}},a_{2,\mathrm{in}},b_{+,\mathrm{in}%
},b_{-,\mathrm{in}}\right) ^{T}$, $V_{\mathrm{noise}}^{\prime }\left(
t\right) =\left( a_{1,\mathrm{vac}},a_{2,\mathrm{vac}},b_{+,\mathrm{th}%
},b_{-,\mathrm{th}}\right) ^{T}$, and%
\begin{equation}
M^{\prime }=\left(
\begin{array}{cccc}
\frac{\kappa }{2}+i\Delta _{1} & iJ_{c} & i\frac{G_{1}e^{i\phi _{1}}}{\sqrt{2%
}} & i\frac{G_{1}e^{i\phi _{1}}}{\sqrt{2}} \\
iJ_{c} & \frac{\kappa }{2}+i\Delta _{2} & i\frac{G_{2}e^{i\phi _{2}}}{\sqrt{2%
}} & -i\frac{G_{2}e^{i\phi _{2}}}{\sqrt{2}} \\
i\frac{G_{1}e^{-i\phi _{1}}}{\sqrt{2}} & i\frac{G_{2}e^{-i\phi _{2}}}{\sqrt{2%
}} & \frac{\gamma }{2}+i\omega _{+} & 0 \\
i\frac{G_{1}e^{-i\phi _{1}}}{\sqrt{2}} & -i\frac{G_{2}e^{-i\phi _{2}}}{\sqrt{%
2}} & 0 & \frac{\gamma }{2}+i\omega _{-}%
\end{array}%
\right) ,
\end{equation}
where
\begin{equation}
b_{\pm ,\mathrm{in}}=\frac{1}{\sqrt{2}}\left( b_{1,\mathrm{in}}\pm
b_{2,\mathrm{in}}\right),
\end{equation}%
\begin{equation}
b_{\pm ,\mathrm{th}}=\frac{1}{\sqrt{2}}\left( b_{1,\mathrm{th}}\pm
b_{2,\mathrm{th}}\right).
\end{equation}%
The scattering probabilities $S_{ij}\left( \omega \right)$ between modes $j$ and $i$ ($i,j=1,2,+,-$, for $a_1,a_2,b_+,b_-$) can be obtained by following the standard method as shown in the last paragraph of subsection~\ref{OMP}.

The scattering probabilities $S_{ij}(\omega)$ between different modes are shown as functions of the frequency $(\omega-\omega_m)/\gamma$ in Fig.~\ref{fig5} for (a) $\Phi=\pi/2$ and (b) $\Phi=3\pi/2$. 
We have $S_{++}\left( \omega \right)\approx 1$ and $S_{+1}\left( \omega \right)\approx S_{+2}\left( \omega \right)\approx S_{+-}\left( \omega \right)\approx 0$ for $\omega=\omega_m-J_m$. This means that the mechanical mode $b_+$ is decoupled from the system for large detuning, and only mechanical mode $b_-$ is coupled to the two optical modes for resonant photon transport.
Similarly, we have $S_{--}\left( \omega \right)\approx 1$ and $S_{-1}\left( \omega \right)\approx S_{-2}\left( \omega \right)\approx S_{-+}\left( \omega \right)\approx 0$ for $\omega=\omega_m+J_m$. That is to say the mechanical mode $b_-$ is decoupled from the system for large detuning, and only mechanical mode $b_+$ is coupled to the two optical modes resonantly.

The probability flow charts are shown in Figs.~\ref{fig5}(c)-\ref{fig5}(f) for $\omega=\omega_m\pm J_m$ and $\Phi=\pi/2$ or $\Phi=3\pi/2$.
It is clear that we can realize nonreciprocal routing with optical mode $a_1$ as transmitter, $a_2$ as receiver, and the normal mechanical modes $b_{\pm}$ as two terminals, as shown in Figs.~\ref{fig5}(c) and \ref{fig5}(f).
We can also realize nonreciprocal routing with $a_1$ as receiver, $a_2$ as transmitter, and the normal mechanical modes $b_{\pm}$ as two terminals [see Figs.~\ref{fig5}(d) and \ref{fig5}(e)].
Different from the proposal based on mechanical modes $b_1$ and $b_2$ shown in Fig.~\ref{fig2}, where the signals are transport to $b_1$ and $b_2$ simultaneously, we can transport signals to only one of the terminal ($b_+$ or $b_-$) with the other one decoupled from the optical modes. Moreover, the signals can transport between the two mechanical modes $b_1$ and $b_2$ with 25 percent probability, while there is no phonon transport between the normal mechanical modes ($b_{\pm}$).


\section{Conclusions}

In summary, we have introduced a nonreciprocal router composing of a transmitter, a receiver, and two terminals. 
We have also proposed a scheme to realize the nonreciprocal router based on an optomechanical plaquette consisting of two optical and two mechanical modes, with one optical mode as transmitter, one optical mode as receiver, and two mechanical modes as terminals.
We demonstrated that the transport direction of the signals in the router can be steered by the synthetic flux induced by the external driving fields, and the frequency of the input signals.
We shew that the nonreciprocal router based on optomechanical plaquette are robust against the experimental imperfectness and within the reach of current experimental conditions. 
Our work lays the foundation for useful applications of nonreciprocal routers in quantum network and quantum secure communication.

\vskip 2pc \leftline{\bf Acknowledgement}

This work is supported by the National Natural Science Foundation of China (Grants No.~12064010 and  No.~12247105),
Natural Science Foundation of Hunan Province of China (Grant
No.~2021JJ20036), and the Science and Technology Innovation Program of Hunan Province (Grant No.~2022RC1203).

\appendix*
\section{Analytical expressions of the elements in scattering matrix}
The elements of the matrix $U(\omega)$ in Eq.~(\ref{UU}) are given by
\begin{widetext}
\begin{eqnarray}
U_{11}\left( \omega \right)  &=&\frac{\kappa _{e}}{Z}\left\{ \left[ \left(
\frac{\gamma }{2}+i\omega _{1}-i\omega \right) \left( \frac{\gamma }{2}%
+i\omega _{2}-i\omega \right) +J_{m}^{2}\right] \left( \frac{\kappa }{2}%
+i\Delta _{2}-i\omega \right) +G_{2}^{2}\left( \frac{\gamma }{2}+i\Delta
_{1}-i\omega \right) \right\} -1, \label{A1} \\
U_{12}\left( \omega \right)  &=&-\frac{\kappa _{e}}{Z}\left[ iJ_{c}\left(
\frac{\gamma }{2}+i\omega _{1}-i\omega \right) \left( \frac{\gamma }{2}%
+i\omega _{2}-i\omega \right) -iJ_{m}G_{1}G_{2}e^{i\left( \phi _{1}-\phi
_{2}\right) }+iJ_{c}J_{m}^{2}\right] , \\
U_{13}\left( \omega \right)  &=&\frac{\sqrt{\kappa _{e}\gamma _{e}}}{Z}\left[
-iG_{2}^{2}G_{1}e^{i\phi _{1}}+iJ_{m}J_{c}G_{2}e^{i\phi _{2}}-iG_{1}e^{i\phi
_{1}}\left( \frac{\kappa }{2}+i\Delta _{2}-i\omega \right) \left( \frac{%
\gamma }{2}+i\omega _{2}-i\omega \right) \right] , \\
U_{14}\left( \omega \right)  &=&-\frac{\sqrt{\kappa _{e}\gamma _{e}}}{Z}%
\left[ J_{c}G_{2}e^{i\phi _{2}}\left( \frac{\gamma }{2}+i\omega _{1}-i\omega
\right) +J_{m}G_{1}e^{i\phi _{1}}\left( \frac{\kappa }{2}+i\Delta
_{2}-i\omega \right) \right] ,
\end{eqnarray}%
\begin{eqnarray}
U_{21}\left( \omega \right)  &=&-\frac{\kappa _{e}}{Z}\left[ iJ_{c}\left(
\frac{\gamma }{2}+i\omega _{1}-i\omega \right) \left( \frac{\gamma }{2}%
+i\omega _{2}-i\omega \right) -iJ_{m}G_{1}G_{2}e^{-i\left( \phi _{1}-\phi
_{2}\right) }+iJ_{c}J_{m}^{2}\right] , \\
U_{22}\left( \omega \right)  &=&\frac{\kappa _{e}}{Z}\left\{ \left[ \left(
\frac{\gamma }{2}+i\omega _{1}-i\omega \right) \left( \frac{\gamma }{2}%
+i\omega _{2}-i\omega \right) +J_{m}^{2}\right] \left( \frac{\kappa }{2}%
+i\Delta _{1}-i\omega \right) +G_{1}^{2}\left( \frac{\gamma }{2}+i\omega
_{2}-i\omega \right) \right\} -1, \\
U_{23}\left( \omega \right)  &=&-\frac{\sqrt{\kappa _{e}\gamma _{e}}}{Z}%
\left[ J_{m}G_{2}e^{i\phi _{2}}\left( \frac{\kappa }{2}+i\Delta _{1}-i\omega
\right) +J_{c}G_{1}e^{i\phi _{1}}\left( \frac{\gamma }{2}+i\omega
_{2}-i\omega \right) \right] , \\
U_{24}\left( \omega \right)  &=&\frac{\sqrt{\kappa _{e}\gamma _{e}}}{Z}\left[
-iG_{2}G_{1}^{2}e^{i\phi _{2}}+iJ_{c}J_{m}G_{1}e^{i\phi _{1}}-iG_{2}e^{i\phi
_{2}}\left( \frac{\kappa }{2}+i\Delta _{1}-i\omega \right) \left( \frac{%
\gamma }{2}+i\omega _{1}-i\omega \right) \right] ,
\end{eqnarray}%
\begin{eqnarray}
U_{31}\left( \omega \right)  &=&\frac{\sqrt{\kappa _{e}\gamma _{e}}}{Z}\left[
-iG_{2}^{2}G_{1}e^{-i\phi _{1}}+iJ_{c}J_{m}G_{2}e^{-i\phi
_{2}}-iG_{1}e^{-i\phi _{1}}\left( \frac{\kappa }{2}+i\Delta _{2}-i\omega
\right) \left( \frac{\gamma }{2}+i\omega _{2}-i\omega \right) \right] , \\
U_{32}\left( \omega \right)  &=&-\frac{\sqrt{\kappa _{e}\gamma _{e}}}{Z}%
\left[ J_{m}G_{2}e^{-i\phi _{2}}\left( \frac{\kappa }{2}+i\Delta
_{1}-i\omega \right) +J_{c}G_{1}e^{-i\phi _{1}}\left( \frac{\gamma }{2}%
+i\Delta _{2}-i\omega \right) \right] , \\
U_{33}\left( \omega \right)  &=&\frac{\gamma _{e}}{Z}\left\{ \left[ \left(
\frac{\kappa }{2}+i\Delta _{1}-i\omega \right) \left( \frac{\kappa }{2}%
+i\Delta _{2}-i\omega \right) +J_{c}^{2}\right] \left( \frac{\gamma }{2}%
+i\omega _{2}-i\omega \right) +G_{2}^{2}\left( \frac{\kappa }{2}+i\Delta
_{1}-i\omega \right) \right\} -1, \\
U_{34}\left( \omega \right)  &=&-\frac{\gamma _{e}}{Z}\left[ iJ_{m}\left(
\frac{\kappa }{2}+i\Delta _{1}-i\omega \right) \left( \frac{\kappa }{2}%
+i\Delta _{2}-i\omega \right) -iJ_{c}G_{1}G_{2}e^{-i\left( \phi _{1}-\phi
_{2}\right) }+iJ_{c}^{2}J_{m}\right] ,
\end{eqnarray}%
\begin{eqnarray}
U_{41}\left( \omega \right)  &=&-\frac{\sqrt{\kappa _{e}\gamma _{e}}}{Z}%
\left[ J_{c}G_{2}e^{-i\phi _{2}}\left( \frac{\gamma }{2}+i\omega
_{1}-i\omega \right) +J_{m}G_{1}e^{-i\phi _{1}}\left( \frac{\kappa }{2}%
+i\Delta _{2}-i\omega \right) \right] , \\
U_{42}\left( \omega \right)  &=&\frac{\sqrt{\kappa _{e}\gamma _{e}}}{Z}\left[
-iG_{2}e^{-i\phi _{2}}\left( \frac{\kappa }{2}+i\Delta _{1}-i\omega \right)
\left( \frac{\gamma }{2}+i\omega _{1}-i\omega \right)
+iJ_{c}J_{m}G_{1}e^{-i\phi _{1}}-iG_{1}^{2}G_{2}e^{-i\phi _{2}}\right] , \\
U_{43}\left( \omega \right)  &=&-\frac{\gamma _{e}}{Z}\left[ iJ_{m}\left(
\frac{\kappa }{2}+i\Delta _{1}-i\omega \right) \left( \frac{\kappa }{2}%
+i\Delta _{2}-i\omega \right) -iG_{1}G_{2}J_{c}e^{i\left( \phi _{1}-\phi
_{2}\right) }+iJ_{c}^{2}J_{m}\right] , \\
U_{44}\left( \omega \right)  &=&\frac{\gamma _{e}}{Z}\left\{ \left[ \left(
\frac{\kappa }{2}+i\Delta _{1}-i\omega \right) \left( \frac{\kappa }{2}%
+i\Delta _{2}-i\omega \right) +J_{c}^{2}\right] \left( \frac{\gamma }{2}%
+i\omega _{1}-i\omega \right) +G_{1}^{2}\left( \frac{\kappa }{2}+i\Delta
_{2}-i\omega \right) \right\} -1, \label{A16}
\end{eqnarray}%
and%
\begin{eqnarray}
Z &=&\left( \frac{\kappa }{2}+i\Delta _{1}-i\omega \right) \left( \frac{%
\kappa }{2}+i\Delta _{2}-i\omega \right) \left( \frac{\gamma }{2}+i\omega
_{1}-i\omega \right) \left( \frac{\gamma }{2}+i\omega _{2}-i\omega \right)
\nonumber \\
&&+G_{1}^{2}\left( \frac{\kappa }{2}+i\Delta _{2}-i\omega \right) \left(
\frac{\gamma }{2}+i\omega _{2}-i\omega \right) +G_{2}^{2}\left( \frac{\kappa
}{2}+i\Delta _{1}-i\omega \right) \left( \frac{\gamma }{2}+i\omega
_{1}-i\omega \right)   \nonumber \\
&&+J_{c}^{2}\left( \frac{\gamma }{2}+i\omega _{1}-i\omega \right) \left(
\frac{\gamma }{2}+i\omega _{2}-i\omega \right) +J_{m}^{2}\left( \frac{\kappa
}{2}+i\Delta _{1}-i\omega \right) \left( \frac{\kappa }{2}+i\Delta
_{2}-i\omega \right)   \nonumber \\
&&+J_{c}^{2}J_{m}^{2}+G_{1}^{2}G_{2}^{2}-2J_{c}J_{m}G_{1}G_{2}\cos \left(
\phi _{1}-\phi _{2}\right) .
\end{eqnarray}
\end{widetext}
\bibliography{ref}

\end{document}